\documentclass{article}
\usepackage{frascatiphys_R}
\begin{document}
\title{PARTICLE PHYSICS AND COSMOLOGY}
\author{Juan Garc\'\i a-Bellido\\
\em Departamento de F\'\i sica Te\'orica C-XI, \\
\em Universidad Aut\'onoma de Madrid, Cantoblanco 28049 Madrid, Spain}
\maketitle
\baselineskip=11.6pt
\begin{abstract}
In this talk I will review the present status of inflationary cosmology
and its emergence as the basic paradigm behind the Standard Cosmological
Model, with parameters determined today at better than 10\% level from
CMB and LSS observations. 
\end{abstract}
\baselineskip=14pt
\section{Introduction}
In this short review I will outline the reasons why the inflationary
paradigm~\cite{inflation,textbooks} has become the backbone of the present
Standard Cosmological Model. It gives a framework in which to pose all
the basic cosmological questions: what is the shape and size of the
universe, what is the matter and energy content of the universe, where
did all this matter come from, what is the fate of the universe, etc. I
will describe the basic predictions that inflation makes, most of which
have been confirmed only recently, while some are imminent, and then
explore the recent theoretical developments on the theory of reheating
after inflation and cosmological particle production, which might allow
us to answer some of the above questions in the future.

Although the simplest slow-roll inflation model is consistent with the
host of high precision cosmological observations of the last few years,
we still do not know what the true nature of the inflaton is: although
there are many possible realizations, there is no unique particle
physics model of inflation. Furthermore, we even ignore the energy scale
at which this extraordinary phenomenon occurred in the early universe;
it could be associated with a GUT theory or even with the EW theory, at
much lower energies.

\section{Basic Predictions}
Inflation is an extremely simple idea based on the early universe
dominance of a vacuum energy density associated with a hypothetical
scalar field called the inflaton. Its nature is not known: whether it is
a fundamental scalar field or a composite one, or something else
altogether. However, one can always use an effective description in
terms of a scalar field with an effective potential driving the
quasi-exponential expansion of the universe. This basic scenario gives
several detailed fundamental predictions: a flat universe with nearly
scale-invariant adiabatic density perturbations with Gaussian initial
conditions.

\subsection{A flat and homogeneous background}
Inflation explains why our local patch of the universe is spatially
flat, i.e.  Euclidean. Inflation does, provides an approximately
constant energy density that induces a tremendous expansion of the
universe. Thus, an initially curved three-space will quickly become
locally indistinguishable from a ``flat'' hypersurface. Moreover, this
same mechanism explains why we see no ripples, i.e. no large
inhomogeneities, in the space-time fabric, e.g. as large anisotropies in
the temperature field of the cosmic microwave background when we look in
different directions.  The expansion during inflation erases any prior
inhomogeneities. These two are very robust predictions of inflation, and
have been confirmed to high precision by the detailed observations of
the CMB, first by COBE (1992) for the large scale homogeneity, to one
part in $10^5$, and recently by BOOMERanG~\cite{deBernardis:2000gy} and
MAXIMA~\cite{Hanany:2000qf}, for the spatial flatness, to better than
10\%.

\subsection{Cosmological perturbations}
Inflation also predicts that on top of this homogeneous and flat
space-time background, there should be a whole spectrum of cosmological
perturbations, both scalar (density perturbations) and tensor
(gravitational waves). These arise as quantum fluctuations of the metric
and the scalar field during inflation, and are responsible for a scale
invariant spectrum of temperature and polarization fluctuations in the
CMB, as well as for a stochastic background of gravitational waves. The
temperature fluctuations were first discovered by COBE and later
confirmed by a host of ground and balloon-borne experiments, while the
polarization anisotropies have only recently been discovered by the CMB
experiment DASI~\cite{pol}. Both observations seem to agree with a
nearly scale invariant spectrum of perturbations. It is expected that
the stochastic background of gravitational waves produced during
inflation could be detected with the next generation of gravitational
waves interferometers (e.g. LISA), or indirectly by measuring the power
spectra of polarization anisotropies in the CMB by the future Planck
satellite~\cite{Planck}.

Inflation makes very specific predictions as to the nature of the scalar
perturbations. In the case of a single field evolving during inflation,
the perturbations are predicted to be adiabatic, i.e. all components of
the matter and radiation fluid should have equal density contrasts, due
to their common origin. As the plasma (mainly baryons) falls in the
potential wells of the metric fluctuations, it starts a series of
acoustic compressions and rarefactions due to the opposing forces of
gravitational collapse and radiation pressure.  Adiabatic fluctuations
give a very concrete prediction for the position and height of the
acoustic peaks induced in the angular power spectrum of temperature and
polarization anisotropies. This has been confirmed to better than 1\% by
the recent observations, and constitutes one of the most important
signatures in favor of inflation, ruling out a hypothetically large
contribution from active perturbations like those produced by cosmic
strings or other topological defects.

Furthermore, the quantum origin of metric fluctuations generated during
inflation allows one to make a strong prediction on the statistics of
those perturbations: inflation stretches the vacuum state fluctuations
to cosmological scales, and gives rise to a Gaussian random field, and
thus metric fluctuations are in principle characterized solely by their
two-point correlation function. Deviations from Gaussianty would
indicate a different origin of fluctuations, e.g. from cosmic defects.
Recent observations by BOOMERanG in the CMB and by gravitational lensing
of LSS indicate that the non-Gaussian component of the temperature
fluctuations and the matter distribution on large scales is strongly
constrained, and consistent with foregrounds (in the case of CMB) and
with non-linear gravitational collapse (in the case of LSS).

Of course, in order to really confirm the idea of inflation one needs to
find cosmological observables that will allow us to correlate the scalar
and the tensor metric fluctuations with one another, since they both
arise from the same inflaton field fluctuations. This is a daunting
task, given that we ignore the absolute scale of inflation, and thus the
amplitude of tensor fluctuations (only sensitive to the total energy
density). The smoking gun could be the observation of a stochastic
background of gravitational waves by the future gravitational wave
interferometers and the subsequent confirmation by detection of the curl
component of the polarization anisotropies of the CMB. Although the
gradient component has recently been detected by DASI, we may have to
wait for Planck for the detection of the curl component.

\section{Recent Cosmological Observations}

Cosmology has become in the last few years a phenomenological science,
where the basic theory (based on the hot Big Bang model after inflation)
is being confronted with a host of cosmological observations, from the
microwave background to the large scale distribution of matter, from the
determination of light element abundances to the detection of distant
supernovae that reflect the acceleration of the universe, etc. I will
briefly review here the recent observations that have been used to
define a consistent cosmological standard model.

\subsection{Cosmic Microwave Background}

The most important cosmological phenomenon from which one can extract
essentially all cosmological parameters is the microwave background and,
in particular, the last scattering surface temperature and polarization
anisotropies. Since they were discovered by COBE in 1992, the
temperature anisotropies have lived to their promise. They allow us to
determine a whole set of both background (0-th order) and perturbation
(1st-order) parameters -- the geometry, topology and evolution of
space-time, its matter and energy content, as well as the amplitude and
tilt of the scalar and tensor fluctuation power spectra -- in some cases
to better than 10\% accuracy.

At present, the forerunners of CMB experiments are BOOMERanG and MAXIMA
(balloons), and DASI, VSA and CBI (ground based interferometers).
Together they have allowed cosmologists to determine the angular power
spectrum of temperature fluctuations down to multipoles 1000 and 3000,
respectively, and therefore provided a measurement of the positions and
heigths of at least 3 to 7 acoustic peaks. A combined analysis of the
different CMB experiments yields convincing evidence that the universe
is flat, with $|\Omega_K| = |1-\Omega_{\rm tot}| < 0.05$ at 95\% c.l.;
full of dark energy, $\Omega_\Lambda = 0.66\pm0.06$, and dark matter,
$\Omega_m = 0.33\pm0.07$, with about 5\% of baryons, $\Omega_b =
0.05\pm0.01$; and expanding at a rate $H_0 = 68\pm7$ km/s/Mpc, all
values given with $1\sigma$ errors, see Table~\ref{table}.  The spectrum
of primordial perturbations that gave rise to the observed CMB
anisotropies is nearly scale-invariant, $n_s = 1.02\pm0.06$, adiabatic
and Gaussian distributed. This set of parameters already constitutes the
basis for a truly Standard Model of Cosmology, based on the Big Bang
theory and the inflationary paradigm.  Note that both the baryon content
and the rate of expansion determinations with CMB data alone are in
excellent agreement with direct determinations from BBN light element
abundances~\cite{O'Meara:2000dh} and HST
Cepheids~\cite{Freedman:2000cf}, respectively.

\begin{table}[t]
\centering
\caption{ \it Estimates of the cosmological parameters that 
characterize a minimal adiabatic inflation-based model. 
From Ref.~\cite{CBI}.
}
\vskip 0.1 in
\begin{tabular}{|l|c|c|c|c|} \hline
        Priors  &  CMB & CMB+LSS & CMB+LSS+SN & CMB+LSS+SN+HST \\
\hline \hline
 $\Omega_{\rm tot}$   & $1.05^{+0.05}_{-0.05}$ & $1.03^{+0.03}_{-0.04}$ 
& $1.00^{+0.03}_{-0.02}$ & $1.00^{+0.02}_{-0.02}$ \\[4pt]
  $n_s$ & $1.02^{+0.06}_{-0.07}$ & $1.00^{+0.06}_{-0.06}$ 
& $1.03^{+0.06}_{-0.06}$ & $1.04^{+0.05}_{-0.06}$ \\[4pt]
  $\Omega_b\,h^2$  & $0.023^{+0.003}_{-0.003}$ & $0.023^{+0.003}_{-0.003}$ & $0.024^{+0.003}_{-0.003}$ & $0.024^{+0.002}_{-0.003}$  \\[4pt]
  $\Omega_{\rm cdm}h^2$  & $0.13^{+0.03}_{-0.02}$ & $0.12^{+0.02}_{-0.02}$ & $0.12^{+0.02}_{-0.02}$ & $0.12^{+0.01}_{-0.01}$ \\[4pt]
  $\Omega_\Lambda$  & $0.54^{+0.12}_{-0.13}$ & $0.61^{+0.09}_{-0.10}$ & $0.69^{+0.04}_{-0.06}$ & $0.70^{+0.02}_{-0.03}$ \\[4pt]
  $\Omega_m$  & $0.52^{+0.15}_{-0.15}$ & $0.42^{+0.12}_{-0.12}$ & $0.32^{+0.06}_{-0.06}$ & $0.30^{+0.02}_{-0.02}$ \\[4pt]
  $\Omega_b$  & $0.080^{+0.023}_{-0.023}$ & $0.067^{+0.018}_{-0.018}$ & $0.052^{+0.011}_{-0.011}$ & $0.049^{+0.004}_{-0.004}$ \\[4pt]
  $h$  & $0.55^{+0.09}_{-0.09}$ & $0.60^{+0.09}_{-0.09}$ & $0.68^{+0.06}_{-0.06}$ & $0.69^{+0.02}_{-0.02}$ \\[4pt]
  Age  & $15.0^{+1.1}_{-1.1}$ & $14.7^{+1.2}_{-1.2}$ & $13.8^{+0.9}_{-0.9}$ & $13.6^{+0.2}_{-0.2}$ \\[4pt]
  $\tau_c$  & $0.16^{+0.18}_{-0.13}$ & $0.09^{+0.12}_{-0.07}$ &
$0.13^{+0.14}_{-0.10}$ & $0.13^{+0.13}_{-0.10}$ \\
\hline
\end{tabular}\\[4pt]
The age of the Universe is in Gyr, and the rate of expansion in \\
units of 100 km/s/Mpc. All values quoted with $1\sigma$ errors.
\label{table}
\end{table}

In the near future, a new satellite experiment, the Microwave Anisotropy
Probe (MAP)~\cite{MAP}, will provide a full-sky map of temperature (and
possibly also polarization) anisotropies and determine the first 2000
multipoles with unprecedented accuracy. When combined with LSS and SN
measurements, it promises to allow the determination of most cosmological
parameters with errors down to the few\% level.

Moreover, with the recent detection of microwave background polarization
anisotropies by DASI~\cite{pol}, confirming the basic paradigm behind
the Cosmological Standard Model, a new window opens which will allow yet
a better determination of cosmological parameters, thanks to the very
sensitive (0.1$\mu$K) and high resolution (4 arcmin) future satellite
experiment Planck~\cite{Planck}. In principle, Planck should be able to
detect not only the gradient component of the CMB polarization, but also
the curl component, if the scale of inflation is high enough. In that
case, there might be a chance to really test inflation through
cross-checks between the scalar and tensor spectra of fluctuations,
which are predicted to arise from the same inflaton potential.

The observed positions of the acoustic peaks of the CMB anisotropies
strongly favor purely adiabatic density perturbations, as arise in the
simplest single-scalar-field models of inflation. These models also
predict a nearly Gaussian spectrum of primordial perturbations. A small
degree of non-gaussianity may arise from self-coupling of the inflaton
field (although it is expected to be very tiny, given the observed small
amplitude of fluctuations), or from two-field models of inflation. Since
the CMB temperature fluctuations probe directly primordial density
perturbations, non-gaussianity in the density field should lead to a
corresponding non-gaussianity in the temperature maps. However, recent
searches for non-Gaussian signatures in the CMB have only given
stringent upper limits, see Ref.~\cite{Polenta:2002}.

One of the most interesting aspects of the present progress in
cosmological observations is that they are beginning to probe the same
parameters or the same features at different time scales in the
evolution of the universe. We have already mentioned the determination
of the baryon content, from BBN (light element abundances) and from the
CMB (acoustic peaks), corresponding to totally different physics and yet
giving essentially the same value within errors. Another example is the
high resolution images of the CMB anisotropies by CBI~\cite{CBI}, which
constitute the first direct detection of the seeds of clusters of
galaxies, the largest gravitationally bound systems in our present
universe. In the near future we will be able to identify and put into
one-to-one correspondence tiny lumps in the CMB with actual clusters
today.

\subsection{Large Scale Structure}

The last decade has seen a tremendous progress in the determination
of the distribution of matter up to very large scales. The present
forerunners are the 2dF Galaxy Redshift Survey~\cite{2dFGRS} and the 
Sloan Digital Sky Survey (SDSS)~\cite{SDSS}. These deep surveys aim
at $10^6$ galaxies and reach redshifts of order 1 for galaxies
and order 5 for quasars. They cover a wide fraction of the sky and
therefore can be used as excellent statistical probes of large scale
structure~\cite{Peacock:2002,Percival:2002}.

The main output of these galaxy surveys is the two-point (and higher)
spatial correlation functions of the matter distribution or,
equivalently, the power spectrum in momentum space. Given a concrete
type of matter, e.g. adiabatic vs. isocurvature, cold vs. hot, etc., the
theory of linear (and non-linear) gravitational collapse gives a very
definite prediction for the measured power spectrum, which can then be
compared with observations. This quantity is very sensitive to various
cosmological parameters, mainly the dark matter content and the baryonic
ratio to dark matter, as well as the universal rate of expansion; on the
other hand, it is mostly insensitive to the cosmological constant since
the latter has only recently (after redshift $z\sim1$) started to become
important for the evolution of the universe, while galaxies and clusters
had already formed by then. Together, 2dFGRS, plus CMB, weak
gravitational lensing and Lyman-$\alpha$ forest data, allow us to
determine the power spectrum with better than 10\% accuracy for $k>0.02
\ h$ Mpc$^{-1}$, which is well fitted by a flat CDM model with
$\Omega_m\,h = 0.20 \pm 0.03$, and a baryon fraction of $\Omega_b /
\Omega_m = 0.15 \pm 0.06$, which together with the HST results give
values of the parameters that are compatible with those obtained with
the CMB, see Table~\ref{table}. It is very reassuring to note that
present parameter determination is robust as we progress from weak
priors to the full cosmological information available, a situation very
different from just a decade ago, where the errors were mostly
systematic and parameters could only be determined with an
order-of-magnitude error. In the very near future such errors will drop
again to the 1\% level, making Cosmology a mature science, with many
independent observations confirming and further constraining previous
measurements of the basic parameters.

An example of such progress appears in the analysis of non-Gaussian
signatures in the primordial spectrum of density perturbations. The
tremendous increase in data due to 2dFGRS and SDSS has allowed
cosmologists to probe the statistics of the matter distribution on very
large scales and infer from it that of the primordial spectrum.
Recently, both groups have reported non-Gaussian signatures (in
particular the first two higher moments, skewness and kurtosis), that
are consistent with gravitational collapse of structure that was
originally Gaussianly distributed~\cite{Verde:2000vr,Szapudi:2001mh}. 
Moreover, weak gravitational lensing also allows an independent
determination of the three-point shear correlation function, and there
has recently been a claim of detection of non-Gaussian signatures in the
VIRMOS-DESCART lensing survey~\cite{Bernardeau:2002}, which is also
consistent with theoretical expectations of gravitational collapse of
Gaussianly distributed initial perturbations.

The recent precise catalogs of the large scale distribution of matter
allows us to determine not only the (collapsing) cold dark matter
content, but also put constraints on the (diffusing) hot dark matter,
since it would erase all structure below a scale that depends on the
free streaming length of the hot dark matter particle. In the case of
relic neutrinos we have extra information because we know precisely
their present energy density, given that neutrinos decoupled when the
universe had a temperature around 0.8 MeV and cooled down ever
since. Their number density today is around 100 neutrinos/cm$^3$. If
neutrinos have a significant mass (above $10^{-3}$ eV, as observations
of neutrino oscillations by SuperKamiokande~\cite{SK} and Sudbury
Neutrino Observatory~\cite{SNO} seem to indicate), then the relic
background of neutrinos is non-relativistic today and could contribute a
large fraction of the critical density, $\Omega_\nu = m_\nu/92\,h^2\,
{\rm eV}\geq 0.001$, see Ref.~\cite{Hu:1997mj}. Using observations of
the Lyman-$\alpha$ forest in absorption spectra of quasars, due to a
distribution of intervening clouds, a limit on the absolute mass of all
species of neutrinos can be obtained~\cite{Croft:1999mm}. Recently, the
2dFGRS team~\cite{Elgaroy:2002bi} have derived a bound on the allowed
amount of hot dark matter, $\Omega_\nu < 0.13\,\Omega_m < 0.05$ (95\%
c.l.), which translates into an upper limit on the total neutrino mass,
$m_{\nu,\ {\rm tot}} < 1.8$ eV, for values of $\Omega_m$ and the Hubble
constant in agreement with CMB and SN observations. This bound improves
several orders of magnitude on the direct experimental limit on the muon
and tau neutrino masses, and is comparable to present experimental
bounds on the electron neutrino mass~\cite{PDG}.

\subsection{Cosmological constant and rate of expansion}

Observations of high redshift supernovae by two independent groups, the
Supernova Cosmology Project~\cite{SNCP}, and the High Redshift Supernova
Team~\cite{HRST}, give strong evidence that the universe is
accelerating, instead of decelerating, today. Although a cosmological
constant is the natural suspect for such a ``crime'', its tiny non-zero
value makes theoretical physicists uneasy~\cite{Weinberg}.  A compromise
could be found by setting the fundamental cosmological constant to zero,
by some yet unknown principle possibly related with quantum gravity, and
allow a super-weakly-coupled homogeneous scalar field to evolve down an
almost flat potential. Such a field would induce an effective
cosmological constant that could in principle account for the present
observations.  The way to distinguish it from a true cosmological
constant would be through its equation of state, since such a type of
smooth background is a perfect fluid but does not satisfy $p=-\rho$
exactly, and thus $w=p/\rho$ also changes with time. There is a proposal
for a satellite called the Supernova / Acceleration Probe
(SNAP)~\cite{SNAP} that will be able to measure the light curves of type
Ia supernovae up to redshift $z\sim2$, thus determining both $\Omega_X$
and $w_X$ with reasonable accuracy, where $X$ stands for this
hypothetical scalar field. For the moment there are only upper bounds,
$w_X < -0.6$ (95\% c.l.)~\cite{PTW}, consistent with a true cosmological
constant, but the SNAP project claims it could determine $\Omega_X$ and
$w_X$ with 5\% precision.

Fortunately, the SN measurements of the acceleration of the universe
give a linear combination of cosmological parameters that is almost
orthogonal, in the plane ($\Omega_m,\ \Omega_\Lambda$), to that of the
curvature of the universe ($1-\Omega_K = \Omega_m + \Omega_\Lambda$) by
CMB measurements and the matter content by LSS data. Therefore, by
combining the information from SNe with that of the CMB and LSS, one can
significantly reduce the errors in both $\Omega_m$ and $\Omega_\Lambda$,
see Table~\ref{table}. It also allows an independent determination of
the rate of expansion of the universe that is perfectly compatible with
the HST data~\cite{Freedman:2000cf}.  This is reflected on the fact that
adding the latter as prior does not affect significantly the mean value
of most cosmological parameters, only the error bars, and can be taken
as an indication that we are indeed on the right track: the Standard
Cosmological Model is essentially correct, we just have to improve the
measurements and reduce the error bars.

\section{Conclusions}

Inflation is nowadays a robust paradigm with a host of
cosmological observations confirming many of its basic predictions:
large scale spatial flatness and homogeneity, as well as an
approximately scale-invariant Gaussian spectrum of adiabatic density
perturbations.

It is possible that in the near future the next generation of CMB
satellites (MAP and Planck) may detect the tensor or gravitational wave
component of the polarization power spectrum, raising the possibility
of really testing inflation through the comparison of the scalar and
tensor components, as well as determining the energy scale of inflation.

\end{document}